\def\RELEASE{0}  %
\def\ANON{0}     %
\def\SQUEEZE{0}  %

\documentclass[sigconf,screen,authorversion,nonacm]{acmart}
\PassOptionsToPackage{hyphens}{url}\usepackage{hyperref}

\usepackage[english]{babel}
\usepackage[noend]{algpseudocode}
\usepackage{algorithmicx}
\usepackage{booktabs}
\usepackage{caption}
\usepackage{comment}
\usepackage[inline]{enumitem}
\usepackage{graphicx}
\usepackage{listings}
\usepackage{multirow}
\usepackage{xcolor}
\usepackage{xspace}
\usepackage{libertine}
\usepackage{microtype}
\usepackage{tikz}
\usepackage{gnuplot-lua-tikz}
\usetikzlibrary{positioning,calc,arrows.meta,fit,backgrounds}

\microtypecontext{spacing=nonfrench}

\hypersetup{
  pdflang={en},
  pdfdisplaydoctitle}

\definecolor[named]{OurPurple}{cmyk}{0.55,1,0,0.15}
\definecolor[named]{OurDarkBlue}{cmyk}{1,0.58,0,0.21}
\hypersetup{
  colorlinks,
  linkcolor=OurPurple,
  citecolor=OurPurple,
  urlcolor=OurDarkBlue,
  filecolor=OurDarkBlue}

\if \SQUEEZE 1
\setlist[itemize]{
  leftmargin=*,
  itemsep=2pt,
  topsep=2pt}

\fi

\def\Snospace~{\S{}}

\if \RELEASE 1
  \def\NOTES{0}
\else
  \def\NOTES{1}
\fi
\if \NOTES 1
  \newcommand{\XXX}[1]{{\color{red}{XXX {#1}}}}
  \newcommand{\antoine}[1]{{\color{teal}{[\textbf{AK:} {#1}]}}}
  \newcommand{\todo}[1]{{\color{blue}{TODO: {#1}}}}
\else
  \newcommand{\XXX}[1]{}
  \newcommand{\antoine}[1]{}
  \newcommand{\todo}[1]{}
\fi

\if \ANON 1
  \newcommand{\sys}{Zap\xspace}
\else
  \newcommand{\sys}{Kugelblitz\xspace}
\fi

\setlist[itemize]{leftmargin=*}
\setlist[enumerate]{leftmargin=*}

\acmConference[]{}{}{}
\renewcommand\footnotetextcopyrightpermission[1]{} %
\setcopyright{none}
\acmDOI{}
\acmISBN{}
\acmPrice{}

\begin{document}
\date{}
\title{\sys: Executable, Cost-Aware Design-Space Exploration for Programmable Packet Pipelines}

\if \ANON 1
	\author{Anonymous Submission \#1604 (Pages: \pageref{page:last_body} body,\pageref{page:last} total)}
	\renewcommand{\shortauthors}{Anonymous Submission \#1604}
\else
	\author[Artem Ageev, Antoine Kaufmann]{
		Artem Ageev
		\and
		Antoine Kaufmann\\
		Max Planck Institute for Software Systems
	}
\fi

\begin{abstract}
	Programmable packet-processing pipelines are a core building block of modern SmartNICs and switches, yet their design requires navigating intertwined trade-offs among program feasibility, hardware cost, and system-level performance.
	Existing approaches rely on proxy metrics such as stage or ALU count, which often mispredict capability and end-to-end behavior.
	We present \textbf{\sys}, a framework for \textbf{executable, cost-aware design-space exploration} of programmable packet pipelines.
	\sys decouples packet-processing programs from pipeline architectures and uses compiler-based feasibility checking to prune designs that cannot support target workloads.
	For feasible architectures, \sys automatically generates synthesizable RTL, enabling synthesis-backed area and timing estimation and cycle-accurate full-system evaluation with real application workloads.
	Using representative programs including NAT, firewalling, and an in-network key--value cache, we show that proxy metrics substantially overestimate capability, that performance rankings change under system-level evaluation, and that the cost of supporting richer workloads is highly non-linear.
\end{abstract}
 
\maketitle

\section{Introduction}

Programmable packet-processing pipelines underpin modern networked systems, from SmartNICs~\cite{nvidia:bluefield3,intel:e810} to high-performance switches~\cite{bosshart:rmt}.
These pipelines enable flexible offloads such as packet classification, stateful filtering, telemetry, and in-network caching, while operating under strict line-rate and latency constraints~\cite{pontarelli:flowblaze,kim:tea,basat:pint,jin:netcache,sapio:switchml}.
Designers meet these demands by assembling pipelines from stages with limited compute, memory, and interconnect resources~\cite{bosshart:rmt}.

Designing such pipelines is inherently a \textbf{capability--per\-for\-mance--cost trade-off}.
Pipeline parameters determine which packet-processing programs can be supported at all, how those programs behave under realistic workloads, and how much silicon area and timing margin the design consumes.
In practice, however, these dimensions are evaluated indirectly.
Architects rely on proxy metrics—such as number of stages, ALUs per stage, or maximum clock frequency—to reason about capability, performance, and cost.
Compilers, in turn, typically target a fixed pipeline and only answer the question of feasibility late in the design cycle~
~\cite{jose:reconfigswitches,sivaraman:packettransactions,gao:cat}.

These abstractions are increasingly brittle.
Pipeline capability depends on structural properties such as dependency depth, state-access ordering, and resource connectivity that proxy metrics fail to capture~\cite{shrivastav:mp5,feng:empower,chen:optimusprime}.
Performance conclusions based on isolated pipeline metrics are often misleading once pipelines interact with queues, DMA engines, host CPUs, and software stacks~\cite{kaufmann:flexnic,eran:nica,li:simbricks}.
Finally, the hardware cost of supporting richer workloads is poorly understood and often grows non-linearly due to required structural changes.

\textbf{We argue that effective pipeline design requires executable, feasibility-first exploration with synthesis-backed cost and system-level performance evaluation.}

We present \textbf{\sys}, a framework that makes this methodology practical.
\sys decouples packet-pro\-cess\-ing programs from pipeline architectures, enabling fast feasibility checking that prunes the architectural design space to pipelines that actually support the workloads of interest.
For feasible designs, \sys automatically generates synthesizable RTL, allowing designers to obtain area and timing from hardware synthesis and to evaluate end-to-end performance via cycle-accurate full-system simulation using SimBricks~\cite{li:simbricks}.

Using a suite of packet-processing programs, including NAT, firewalling, and an in-network key–value cache~\cite{eran:nica}, we show that simple hardware summaries substantially overestimate pipeline capability, that proxy metrics can mispredict end-to-end performance, and that the cost of supporting richer workloads is highly non-linear.
By unifying feasibility checking, synthesis-backed cost estimation, and system-level performance evaluation, \sys enables a principled approach to designing programmable packet-processing pipelines.

\medskip\noindent
This paper makes the following contributions:
\begin{itemize}
	\item \textbf{Executable, cost-aware exploration.}
	      A framework for design-space exploration of programmable packet pipelines using synthesizable hardware and full-system evaluation.

	\item \textbf{Capability-aware feasibility pruning.}
	      Fast feasibility checking that decouples program requirements from pi\-pe\-li\-ne implementations to prune unsupported architectures.

	\item \textbf{Realistic cost and performance evaluation.}
	      Synthesis-backed cost estimation and executable system-level performance evaluation with real application workloads.
\end{itemize}

\noindent
We will release \sys as an open-source artefact on publication.
This work does not raise any ethical issues.
\section{Background \& Motivation}%
\label{sec:bg}

Programmable packet-processing pipelines execute packet logic at line rate using a fixed sequence of stages, each with bounded compute and state access per packet. The defining constraint is \textbf{line rate}: each stage must complete within a single cycle (or fixed per-stage budget) as packets advance through the pipeline in a streaming fashion. As a result, feasibility and cost depend strongly on how program dependencies and state accesses interact with the pipeline's \emph{structure}—stage boundaries, resource composition, and interconnect—rather than on aggregate resource counts alone.

\subsection{The Spatial Pipeline Execution Model}

Pipeline architectures are organized as a sequence of stages separated by registers, where each stage provides bounded compute, state access, and interconnect. Programs are executed by compiling them into a \textbf{runtime configuration} that selects operation bindings, routing paths, and state-access modes for each stage. This configuration serves as the pipeline's machine code and fully determines its behavior.
Crucially, dependencies must respect stage boundaries, and interconnect structure constrains which producer–consumer pairs can be connected within the available stage budget. As a result, compilation is a joint \textbf{binding, routing, and configuration} problem rather than simple instruction scheduling.

\subsection{Why Proxy Metrics Fail}

A common way to compare pipeline architectures is via proxy metrics such as number of stages, number of ALUs per stage, or aggregate memory capacity. These metrics are appealing because they are easy to compute and report. However, they are frequently insufficient for answering the question that matters most to a designer: \textbf{Can my workload run on this architecture at line rate?}

Feasibility is not determined by aggregate resource counts alone. It depends on \emph{where} compute sits relative to state access, \emph{how} values can be routed, and whether the architecture supports the specific operation types and access patterns induced by the program. As a result, pipelines with identical stage and ALU counts may differ substantially in feasibility due to interconnect topology, resource placement, or supported operations. This mismatch is amplified by modern workloads, where even simple packet functions include multi-step dependencies and state access.

Proxy metrics are also poor predictors of hardware cost. For example, in our synthesized baselines, a fixed firewall pipeline with 15 ALUs is slightly smaller (2.55\,mm$^2$) than a fixed NAT pipeline with only 5 ALUs (2.64\,mm$^2$), indicating that non-ALU structure can dominate area.

\subsection{Cost of Capability Is Structural}

Even when a program is feasible, supporting it may require architectural features whose cost is not captured by simple proxies. Examples include richer interconnect (to route more values per stage), additional or wider functional units (to support specific operations), and more flexible state-access blocks (to serve multiple accesses or more complex lookup semantics). These features incur area and timing costs that are tightly coupled to the pipeline's organization, motivating a cost-aware exploration approach grounded in realizable hardware rather than abstract models.

Small capability changes can induce large, non-linear cost and timing effects. For example, in an RMT-like pipeline, adding multiplication support via a one-line DSL change increased synthesized area from 3.79 to 5.34\,mm$^2$ and worsened timing, requiring architectural retuning (e.g., ALU pipelining) to recover target frequency.

\subsection{System-Level Performance Requires End-to-End Evaluation}

Finally, pipeline microbenchmarks alone are often insufficient to predict end-to-end performance. Packet-processing pipelines operate inside a larger system: NIC queues, DMA engines, host CPUs, kernel networking stacks, and application software all contribute to throughput and latency. Architectural choices that appear equivalent at the pipeline level can diverge once integrated into a realistic system context due to backpressure, queueing dynamics, interrupt and scheduling behavior, or application-level bottlenecks.

For this reason, a meaningful evaluation of pipeline architectures—especially for application-facing acceleration—must include \textbf{executable, end-to-end execution} in a system environment with real software stacks and realistic network interaction. Without this, conclusions risk being artifacts of idealized assumptions rather than properties of deployed systems.

\begin{figure*}[h]%
  \includegraphics[scale=0.75]{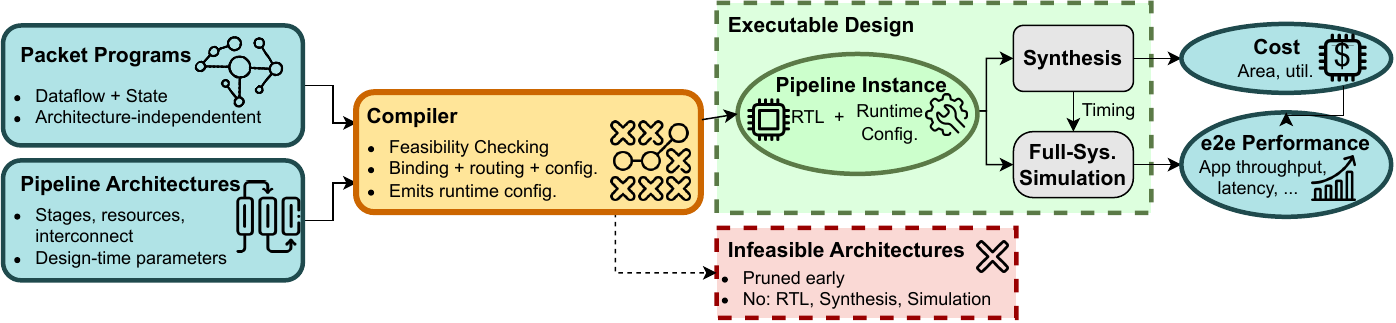}%
  \caption{Feasibility-first executable exploration: compile program+architecture, prune infeasible designs, then use the same generated hardware for synthesis-backed cost and full-system evaluation in simulation.}
  \label{fig:workflow}
\end{figure*}

\section{Problem \& Design Goals}
\label{sec:problem}

We now formalize the programmable packet-pipeline design problem in terms of feasibility, cost, and system-level performance, and derive the design goals for systematic exploration.
\autoref{fig:workflow} illustrates our feasibility-first, executable exploration workflow that underlies this formulation.

\paragraph{Relation to design-space exploration.}
General-purpose de\-sign-space exploration (DSE) and auto-tuning methods assume
(i) feasibility is implicit or inexpensive to test and (ii) most sampled points can be evaluated.
Reconfigurable packet pipelines violate both assumptions: many architecture-pro\-gram pairs are infeasible,
and evaluation is only meaningful after compiliation.
\sys{} therefore exposes feasibility as a first-class predicate and produces an executable artifact
for each feasible design point, enabling DSE methods to operate over a space with pervasive infeasibility.

\subsection{Problem Formulation}

We consider the problem of designing programmable packet-processing pipeline architectures that support a target set of packet-processing programs under line-rate constraints.
A pipeline architecture specifies available resources—stages, functional units, memories, and interconnect—and the structural constraints governing packet execution, while a program defines per-packet computation and state access.

Given an architecture and a program, the first question is feasibility: whether the program can be mapped to the architecture at all. In reconfigurable packet pipelines, feasibility is binary—programs either execute at line rate or cannot be implemented correctly. Feasibility depends on structural properties such as dependency depth, state-access ordering, and resource connectivity, and cannot be inferred reliably from aggregate resource counts.

For feasible designs, two additional dimensions matter.
\textbf{Cost} is modeled using synthesis-backed metrics such as silicon area, and is tightly coupled to capability:
supporting additional programs often requires structural changes that induce non-linear increases in area or timing.
\textbf{Performance} is the behavior of the resulting design in a realistic system and depends on interactions with queues, DMA engines, host CPUs, and application workloads.

In summary, the pipeline design problem is to identify architectures that are feasible for the target programs, achieve acceptable cost, and deliver good system-level performance.
Addressing this requires a methodology that explicitly reasons about feasibility, quantifies cost using executable hardware designs, and evaluates performance in realistic systems.

\subsection{Design-Space Exploration Challenges}

Exploring packet pipeline designs in practice is challenging due to the cost and complexity of evaluating candidate architectures. Determining feasibility requires a compiler capable of mapping programs to a specific pipeline, but existing compilers are architecture-specific and must be built or adapted for each design. Without a generic compiler, even basic questions about program compatibility cannot be answered.

Quantifying \textbf{cost} further raises the bar.
Reliable estimates of area and timing require complete hardware implementation and synthesis using standard toolchains; analytical models fail to capture structural effects such as interconnect complexity.%
In practice, synthesis quickly dominates exploration cost; brute-force, synthesis-in-the-inner-loop exploration is impractical without early pruning.

Evaluating \textbf{performance} is equally challenging. Although feasible pipelines are designed to sustain line rate, their end-to-end behavior depends on system-level interactions with queues, DMA engines, host CPUs, and application workloads. Capturing these effects requires executable hardware integrated into a realistic system environment, which is typically impractical for early-stage exploration.

Together, these challenges make systematic exploration of packet pipeline designs prohibitively expensive. Designers are forced to rely on proxy metrics and intuition, and only exploring few hand-crafted designs thereby obscuring trade-offs among feasibility, cost, and system-level performance.

\subsection{Design Goals}

To make packet pipeline design-space exploration practical, our methodology must satisfy several key goals:

First, it must provide \textbf{explicit feasibility checking}. Given a program and an architecture, the methodology should precisely determine whether the program can be mapped to the pipeline, rather than estimating compatibility from proxy metrics. This feasibility check must be sound and sufficiently efficient as an early-stage filter over large design spaces.

Second, the methodology must enable \textbf{cost-aware evaluation} based on realistic hardware implementations. Cost metrics such as area and timing should be derived from executable hardware designs using standard synthesis flows, allowing meaningful comparison across architectural alternatives without requiring manual RTL development.

Third, the methodology must support \textbf{end-to-end system-level performance evaluation}. Because pipeline behavior interacts with surrounding system components, performance must be measured using executable designs in a realistic system context with representative workloads, rather than relying on isolated pipeline metrics.

Finally, the methodology must impose \textbf{low engineering overhead} and scale to realistic pipeline sizes, enabling exploration across architectures without building architecture-specific compilers, hand-written RTL, or custom testbeds.

\section{\sys: Approach \& Overview}
\label{sec:approach}

\begin{figure}[t]
  \centering
  \includegraphics[scale=0.75]{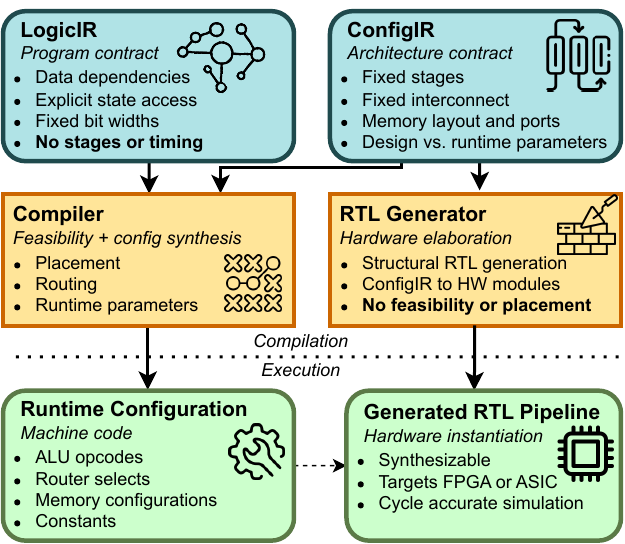}%
  \vspace{-4pt}
  \caption{Abstraction contracts and compilation interface in \sys. The compiler checks feasibility and produces a runtime configuration for the generated RTL.}
  \label{fig:abstractions}%
\end{figure}

\sys is an executable, cost-aware design-space exploration framework for programmable packet-processing pipelines that integrates feasibility checking, structural RTL generation from pipeline descriptions, and validated end-to-end performance evaluation.
\autoref{fig:abstractions} summarizes the abstraction contracts and component interfaces enabling this.
Its architecture is organized to support systematic exploration across the three dimensions identified in Section 3—feasibility, cost, and system-level performance—while keeping engineering effort low. Rather than assuming a fixed datapath, \sys treats pipeline architectures as first-class inputs and enables their evaluation using executable hardware designs.

\subsection{End-to-End Workflow}
Given a packet-processing program and a candidate pipeline architecture, \sys invokes a compiler to determine \textbf{feasibility}, whether the program can be mapped to the architecture under line-rate constraints. Architectures that fail this check are discarded without further evaluation.

For feasible designs, the same compilation step produces the \textbf{architecture-specific runtime configuration} (pipeline machine code) required to run the program on the pipeline.
This configuration assigns runtime parameters to elements, including ALU opcodes, routing selections, and state-access parameters.
Packet-processing pipelines cannot execute programs without these configurations; thus, executable evaluation fundamentally requires a compiler. To date, producing these configurations required building a dedicated compiler tightly coupled to a single pipeline architecture, making executable evaluation across many designs impractical.

\sys removes this barrier by using a single compiler that supports both feasibility checking and runtime configuration generation across a broad space of pipeline architectures. This enables executable evaluation of feasible designs without building architecture-specific toolchains.

For each feasible architecture, \sys automatically generates a complete, synthesizable pipeline RTL implementation. The RTL passes through standard ASIC or FPGA synthesis toolchains to obtain \textbf{area and timing}, to estimate cost and inform simulaiton. The same RTL, together with the compiled runtime configuration, then also integrates into a \textbf{cycle-accurate full-system simulation environment}, enabling evaluation under complete software stacks and application workloads.

With a unified compilation and hardware generation flow, \sys ensures that feasibility, cost, and performance are evaluated on consistent executable designs, enabling systematic, cost-aware exploration of pipeline architectures.

\subsection{Design Principles}
\sys is guided by four design principles that together enable practical, executable design-space exploration.

\paragraph{Explicit feasibility as a first-class concern.}
\sys explicitly checks pro\-gram-architecture compatibility, rather than inferring it from proxy metrics. Early feasibility checks eliminate architectures that cannot support the target programs, avoiding unnecessary hardware generation and evaluation.

\paragraph{Architecture as an input, not an assumption.}
\sys treats pipeline architectures as configurable inputs rather than fixed design choices. This allows systematic exploration across variations in stage structure, resource composition, and interconnect topology, rather than constraining evaluation to a single datapath.

\paragraph{Compilation as an execution enabler.}
The compiler in \sys serves both to determine feasibility and to generate the runtime configuration required to execute programs on candidate architectures. This avoids the need to build architecture-specific compilers for each design and enables executable evaluation across a broad design space.

\paragraph{Executable evaluation of cost and performance.}
\sys evaluates architectures with executable hardware designs. Cost derives from synthesis-backed area and timing, while performance is measured via cycle-accurate full-system simulation with real workloads. This avoids reliance on proxy metrics and ensures that both dimensions reflect structural effects.
\section{Independent Program and Architecture Abstractions}%
\label{sec:design-irs}

\sys relies on explicit representations of packet-pro\-cess\-ing programs (LogicIR) and pipeline architectures (ConfigIR) for feasibility checking and executable evaluation across a broad design space (\autoref{fig:abstractions}). These abstractions capture the structural properties determining program-architecture compatibility, while remaining independent of specific pipeline implementations.
\autoref{tab:knobs} summarizes the compiler-visible knobs in \sys{}.

\begin{table}[t]
	\centering
	\scriptsize
	\setlength{\tabcolsep}{3pt}
	\begin{tabular}{p{1.05cm} p{2.9cm} p{2.9cm}}
		\toprule
		\textbf{IR}                                                & \textbf{Captures} & \textbf{Compiler-instantiated params} \\
		\midrule
		LogicIR                                                    &
		DFG ops + widths; explicit state; packet I/O slices        &
		---                                                                                                                    \\
		ConfigIR                                                   &
		Stages/regs; block library; explicit interconnect topology &
		Router selects; ALU op; const/slice/extend; mem id + R/W                                                               \\
		\bottomrule
	\end{tabular}
	\caption{Compiler-visible knobs. LogicIR describes program structure; ConfigIR describes architectural structure and run-time configuration parameters.}
	\label{tab:knobs}
\end{table}

\subsection{Program Representation}

\sys represents packet-processing programs in an \textbf{arch\-it\-ec\-ture-independent form} that captures the structural constraints relevant to feasibility and execution. Programs comprise two parts: \textbf{persistent state}, which survives across packets, and \textbf{per-packet logic}, which operates on packet fields and state at line rate. Persistent state is declared explicitly as arrays or lookup tables with fixed dimensions (element or key width and capacity), while per-packet logic is expressed as a data-flow graph (detail in \autoref{appendix:logicir}).

The per-packet logic is represented as a \textbf{typed data-flow graph (DFG)} whose nodes correspond to concrete operations and whose edges encode true data dependencies. LogicIR defines a fixed set of operation classes covering bit-width conversions, arithmetic and logical computation, packet input/output, and explicit state access. Each node has well-defined input and output ports, and state-access nodes reference declared state by identifier.

All values in LogicIR are modeled as \textbf{fixed-width bitvectors}. Operations such as slice, merge, and extend explicitly define width transformations, and state declarations fix the widths of stored values and lookup keys. This strict width discipline allows the compiler to reason precisely about data dependencies, compatibility, and feasibility without imposing a sequential execution order. LogicIR intentionally preserves only the partial order induced by data dependencies, exposing available parallelism while abstracting away language-specific syntax and semantic rewrites.

\subsection{Architecture Representation}

\sys represents pipeline architectures with \textbf{ConfigIR}, a structural description of the target datapath. A pipeline is modeled as a directed graph whose nodes correspond to architectural resources, such as registers, ALUs, memories, and routers, and whose edges capture the interconnect topology between them. Input, output, and storage blocks are specified separately and wired explicitly to pipeline ports (\autoref{appendix:configir}).

ConfigIR encodes the staged execution model of reconfigurable packet-processing pipelines directly. \textbf{Register blocks define stage boundaries}, and only values written to stage output registers are visible to the next stage. This structure reflects the one-cycle-per-stage execution model necessary for line-rate execution and constrains placement of dependent operations across stages.

Architectural elements expose a clear separation between \textbf{design-time parameters} (e.g., ALU width, supported opcode set, latency, router fan-in) and \textbf{runtime-configurable parameters} (e.g., opcode selection, router input selection, constant values). This separation defines the degrees of freedom for the compiler and ensures ConfigIR descriptions can be translated directly into realizable hardware. Each architectural element in ConfigIR maps directly to a parameterized hardware module instantiated by the RTL generator.

\begin{figure}[t]
	\centering
	\includegraphics[scale=0.75]{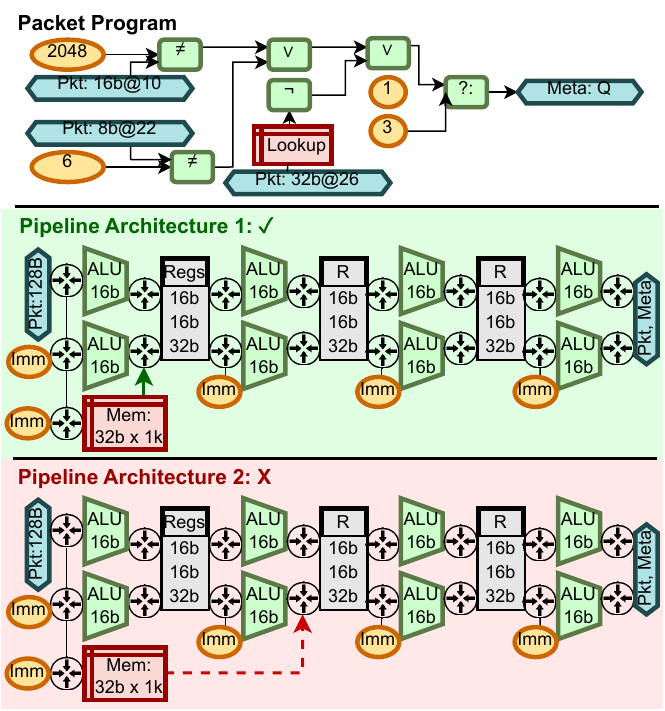}%
	\caption{Program--architecture decoupling example. A packet-processing dataflow graph (top) is infeasible on Pipeline~1 but feasible on Pipeline~2, despite identical stage and ALU counts, due to interconnect difference.}
	\label{fig:decoupling}
\end{figure}

\subsection{Decoupling Program from Architecture}

\sys independently represents programs and architectures. LogicIR captures the structural requirements of a program, while ConfigIR captures the resources and constraints of a pipeline architecture. This decoupling allows evaluation of the same program against many architectures (\autoref{fig:decoupling}), and the same architecture against different workloads.

By separating these concerns, \sys enables feasibility-based pruning of the architectural design space and avoids embedding architectural assumptions into program representations. When feasible, the compiler produces a runtime configuration that drives both synthesized hardware and cycle-accurate simulation, ensuring feasibility, cost, and performance are evaluated on the same executable design.

\subsection{Scope and Limitations}

The abstractions in \sys are designed to capture the structural properties of packet-processing programs and pipeline architectures that determine feasibility and execution at line rate. They do not model full programming-language semantics, semantic rewrites, or control-plane behavior, nor do they capture all microarchitectural timing effects.

These limitations affect \textbf{completeness}, not \textbf{soundness}: when \sys reports a program as feasible, the resulting executable configuration is correct for the modeled pipeline. Extending front-end support or refining the abstractions would broaden coverage without altering the core feasibility and evaluation methodology presented in this paper.
\section{Capability-Aware Feasibility Checking}%
\label{sec:design-compiler}

We now describe how \sys compiles packet-processing programs onto candidate pipeline architectures. The compiler serves two purposes: it determines whether a program is feasible on a given architecture, and, when feasible, produces the executable runtime configuration required to run the program in simulation or hardware. We frame compilation as a feasibility problem under structural constraints, rather than an optimization problem, and outline our constraint solving approach to generality across heterogeneous pipelines.

\subsection{Compiler Role and Outputs}
The \sys compiler consumes our program and architecture representations introduced above and serves two purposes: it determines \textbf{feasibility} (can a LogicIR program map to a given ConfigIR architecture at line-rate?) and, when feasible, produces the \textbf{executable runtime configuration} required to execute the program on that architecture.

Formally, the compiler takes as input a LogicIR program and a ConfigIR hardware configuration and searches for an assignment to the \textbf{runtime configuration parameters of ConfigIR nodes} such that the configured pipeline is equivalent to the LogicIR specification. If such an assignment exists, compilation returns a concrete runtime configuration (“machine code”); otherwise it reports infeasibility.

Compilation proceeds in four stages: (i) deterministic front-end validation of LogicIR/ConfigIR well-formedness, (ii) \textbf{candidate generation} that enumerates compatible hardware implementations for each LogicIR node, (iii) construction of a constraint system encoding placement, routing, and configuration choices, and (iv) SAT solving followed by decoding of the satisfying assignment into a runtime configuration.

\begin{figure}[t]
	\centering
	\includegraphics[scale=0.75]{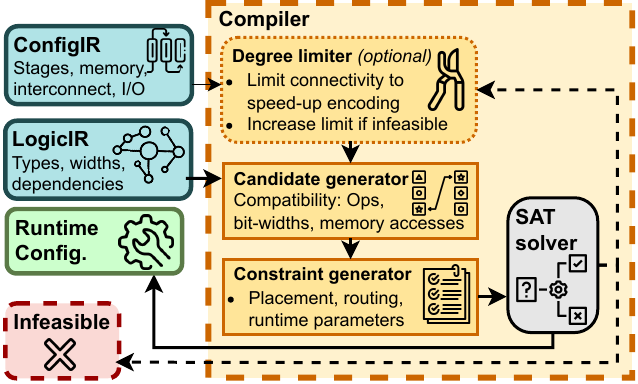}%
	\caption{Feasibility checking as constraint satisfaction: candidate generation plus placement, routing, and configuration constraints. SAT yields a runtime configuration; UNSAT implies infeasibility.}
	\label{fig:mapping}
\end{figure}

\subsection{Feasibility as Constraint Satisfaction}
Compilation in \sys is formulated as a \textbf{feasibility problem}, not an optimization problem.
\autoref{fig:mapping} summarizes how feasibility checking is reduced to constraint satisfaction over placement, routing, and runtime parameters.
Under the execution model of reconfigurable packet-processing pipelines, programs either map and run at line rate or cannot be implemented correctly; there is no meaningful throughput trade-off among feasible mappings.

Feasibility requires jointly deciding three tightly coupled aspects: \textbf{binding} (which hardware node instance implements each LogicIR operation), \textbf{routing} (how values flow through the interconnect to satisfy all LogicIR dependencies), and \textbf{runtime parameter selection} (e.g., ALU opcode choices, router input selections, constant values, slice/extend parameters, and memory access identifiers/modes). These decisions are interdependent: binding choices constrain routing, and routing choices constrain valid runtime parameters.

\autoref{appendix:constraints} formalizes the variables and constraint classes and how a satisfying assignment decodes to runtime configuration.

A key preprocessing step is \textbf{candidate generation}. For each LogicIR node, the compiler computes a finite set of compatible ConfigIR nodes based on operation type, bit-width compatibility, supported opcode sets (for compute), and state-access compatibility (for memory operations). This candidate relation both defines feasibility precisely and substantially reduces the search space before SAT encoding.

\subsection{Constraint Solving Approach}

\sys encodes compilation as a constraint satisfaction problem and solves it using Boolean satisfiability (SAT). The encoding introduces Boolean variables capturing \textbf{placement} (is LogicIR node `l` is implemented by hardware node `h`?), \textbf{routing} (can a produced values reach required consumer inputs through the hardware graph?), and \textbf{runtime configuration choices} (router selections, ALU opcode selections, memory modes, constants, and other node parameters).
We give the complete encoding (including router-selection reachability constraints) in \autoref{appendix:constraints}.

The constraint system enforces a small number of semantically meaningful constraint classes:

\begin{itemize}
	\item \textbf{Placement correctness:} each LogicIR operation is implemented exactly once by a compatible hardware node.
	\item \textbf{Resource consistency:} hardware resources are used consistently with their semantics (e.g., a configured node's runtime parameters match the attributes of the LogicIR operation it implements).
	\item \textbf{Connectivity/routing legality:} for every LogicIR dependency edge, the corresponding value is routable through the ConfigIR interconnect from the producer placement to the consumer placement, consistent with router selections.
	\item \textbf{Packet I/O anchoring:} PacketIn/Out semantics are respected and bound to the corresponding architecture blocks.
\end{itemize}

\noindent
We use a standard SAT solver; solver choice is not fundamental, and we do not rely on solver-specific features.

\subsection{Executable Configuration Generation}
When the SAT instance is satisfiable, the compiler decodes the satisfying assignment into an \textbf{executable runtime configuration}, represented as a per-node assignment of runtime parameters (e.g., router input selections, ALU opcodes, constants, slice/extend parameters, and memory access identifiers/modes). This configuration fully determines the behavior of the configured pipeline for the given program.
\autoref{appendix:constraints} describes decoding the satisfying assignment into per-node machine code.

The runtime configuration is used unchanged to drive both synthesized hardware and cycle-accurate system simulation. As a result, feasibility checking and execution are tightly coupled: the same compilation result that establishes feasibility also defines the concrete pipeline behavior used for cost estimation and end-to-end evaluation.

\subsection{Degree-Limiter}
For scalability, the compiler supports an optional \textbf{degree-limiter} that restricts the neighborhood considered during encoding (e.g., by limiting connectivity considered for candidate placements). This preserves \textbf{soundness}, any satisfying assignment corresponds to a valid implementation, but can sacrifice \textbf{completeness}; UNSAT under the limiter does not necessarily imply infeasibility under the full encoding.
\autoref{appendix:constraints} discusses encoding-time pruning (degree limiter) and its soundness/completeness implications.

\paragraph{UNSAT confirmation.}
Because the degree limiter is sound but incomplete, UNSAT at a limiter value may reflect pruning rather than true infeasibility.
We therefore rerun UNSAT points with a larger limiter (or with the limiter disabled) before using them in plots/tables.

\subsection{Scope, Heuristics, and Limitations}
The compiler solves a pure \textbf{feasibility} problem rather than optimizing among feasible mappings. This choice is justified by the line-rate execution model: if a program maps, it executes at line rate, and if it does not map, it cannot run; feasible assignments yield equivalent throughput.

Finally, the compiler does not perform semantic rewrites or reason about equivalent program formulations; such transformations can layer on top of the feasibility formulation without changing the core compilation and evaluation flow.
\section{HW Generation \& Cost Estimation}%
\label{sec:design-hwgen}

\sys turns feasible pipeline architectures into synthesizable RTL, enabling synthesis-backed area/timing cost metrics and executable system-level performance evaluation from the same generated design.

\subsection{Automatic Hardware Realization}
The \sys hardware generator takes the \textbf{ConfigIR hardware configuration} (pipeline graphs, blocks, and wiring) as input together with a \textbf{target selector} (ASIC, FPGA, or simulation) and emits a \textbf{complete, synthesizable Verilog RTL design}. Hardware generation is a structural translation: ConfigIR nodes instantiate parameterized module templates, and ConfigIR edges realize as Verilog wires connecting module ports.

The generator implements a deterministic mapping from \emph{(ConfigIR node type, design-time parameters, target policy)} to an RTL module exposing a \textbf{runtime configuration interface} consistent with ConfigIR's runtime-configurable fields. Internally, the generator elaborates the ConfigIR graph into a netlist-like intermediate representation (instances + connections) before emitting Verilog.

\subsection{Synthesis-Based Cost Metrics}
The generated RTL then passes through standard ASIC or FPGA synthesis tools to obtain \textbf{area} and \textbf{achievable timing} estimates, which serve as cost metrics for design-space exploration. These metrics are used comparatively rather than as signoff-quality estimates; the RTL captures the full structural complexity of each architecture, including interconnect and explicit state-access components. Synthesis thus reflects cost effects that are difficult to model analytically.

Target-specific policies apply consistently during generation. In particular, state-access nodes bind to target-appropriate memory backends (e.g., ASIC SRAM macros via a memory generator, FPGA BRAM/URAM, or synthesizable Verilog memories for simulation). Synthesis and execution reflect the same architectural structure and state-access placement.

\subsection{Executable Hardware in System Context}
The generated RTL also enables \textbf{cycle-accurate system evaluation}. The generated top-level module translates to ConfigIR's PacketIn/Out blocks, enabling different pipelines to integrate into a system through a well-defined streaming I/O interface.
The compiler-produced runtime configuration drives execution: the simulated software driver feeds the configured values to the generated hardware through the configuration interface to program the runtime-configurable fields of the instantiated modules. Together, the generated RTL and runtime configuration define a fully executable design, unchanged across synthesis and full-system simulation.

\subsection{Scope and Limitations}
The hardware generation flow is intended for \textbf{comparative design-space exploration}, not full signoff physical design. Area and timing estimates depend on the chosen toolchain, technology libraries, and target-specific backend choices, and do not capture layout-, routing-, or power-related effects.

Despite these limitations, the generator produces complete, structurally faithful RTL that instantiates the architectural components and connectivity specified by ConfigIR. This fidelity is sufficient to support synthesis-backed cost comparisons and to execute candidate architectures consistently in system-level evaluation.
\section{E2E Perf. Eval. Methodology}%
\label{sec:design}

\begin{figure}[t]
  \centering
  \includegraphics[scale=0.75]{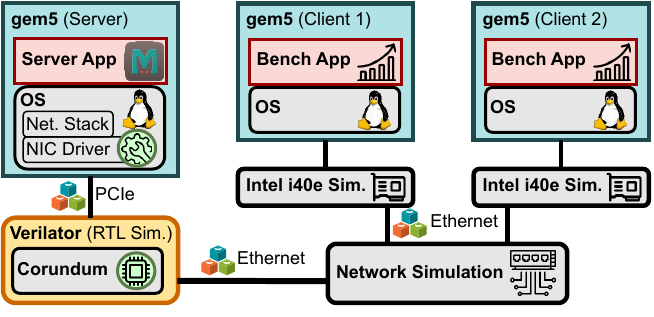}%
  \caption{Example SimBricks full-system simulation of \sys NIC pipeline with hosts; clock from synthesis.}
  \label{fig:simsetup}
\end{figure}

\sys evaluates pipeline architectures in \textbf{complete network systems} (e.g. \autoref{fig:simsetup}). Feasible architectures execute as hardware components in a cycle-accurate full-system simulation~\cite{li:simbricks}, enabling end-to-end performance evaluation with real software stacks and application workloads.

\subsection{Execution in a Full-System Context}

Generated pipeline hardware executes as part of a full system that includes NICs, host CPUs, DMA engines, queues, and a packet-switched network connecting multiple machines. The pipeline operates at the clock frequency obtained from synthesis and is driven by the runtime configuration produced by the compiler. Backpressure, queueing, and contention propagate naturally through the HW-SW boundary.

Unlike pipeline-only simulators or trace-driven evaluations, \sys executes packet pipelines \textbf{in situ}, as components of a networked system where packets originate from and terminate at software running on remote hosts.

\subsection{Software Stack and App Execution}

The simulated hosts run \textbf{unmodified operating systems} and \textbf{unmodified application binaries}. Packet-processing pipelines interact with software through standard NIC interfaces, exercising the full networking stack, memory subsystem, and scheduling behavior. Applications communicate over the simulated network using their native protocols, without synthetic drivers or application shortcuts.
As a result, performance observed in \sys emerge from realistic interactions between hardware pipelines, network behavior, and applications, rather than from isolated microbenchmarks or synthetic traffic generators.

\subsection{Hardware Timing and Interaction}

Pipeline timing derives directly from synthesis results, ensuring execution reflects the structural timing properties of each architecture. Hardware stalls, backpressure, and queue occupancy are modeled cycle-accurately and influence software-visible behavior such as throughput and latency. This enables evaluation of architectural choices whose effects manifest only through system-level interactions.

\subsection{Validation and Scope}

Where possible, we validate \sys simulations against FPGA implementations of the generated hardware. While the evaluation does not model power consumption or physical layout, it provides sufficient fidelity for \textbf{comparative, architecture-level analysis}. By grounding performance measurements in executable hardware and real software execution, \sys enables system-level conclusions beyond proxy metrics.
\section{Evaluation}
\label{sec:eval}

We evaluate \sys along four questions: (1) are our end-to-end performance results trustworthy, i.e., do they match a physical testbed; (2) does the hardware generator produce realistic RTL comparable to hand designs; (3) is the exploration loop practical at realistic pipeline scales in terms of runtime and cost estimation; and (4) does the framework enable rapid, quantitative capability--cost trade-off studies.

\begin{figure}[t]
	\centering
	\begin{tabular}{l rrr}
		\toprule
		\textbf{Architecture} & \textbf{ALUs} & \textbf{Chip Area} & \textbf{LoC} \\
		\midrule
		Flex  10 x 30         & 641           & 8.46\,mm$^2$       & 85           \\

		Flex  3 x 100         & 304           & 3.79\,mm$^2$       & 85           \\
		Flex  3 x 100 (+mul)  & 304           & 8.41\,mm$^2$       & 85           \\

		Flex  30 x 30         & 1921          & 15.12\,mm$^2$      & 85           \\
		Flex  30 x 100        & 3031          & 51.33 mm$^2$       & 85           \\

		Fixed NAT             & 5             & 2.64\,mm$^2$       & 56           \\
		Fixed Firewall        & 15            & 2.55\,mm$^2$       & 111          \\
		\bottomrule
	\end{tabular}
	\caption{Hardware design points used in evaluation (fixed baselines and Flex $m\times n$ reconfigurable pipelines) with synthesized area and DSL size.}
	\label{fig:details}
\end{figure}

\subsection{Setup}
We implemented \sys in Scala. The compiler uses a SAT backend (MiniSat 2.2 in our evaluation), and hardware cost is derived from ASIC synthesis using Synopsys Design Compiler (R-2020.09-SP2) with FreePDK45 libraries. All experiments run on a 2$\times$22-core Intel Xeon Gold 6152 machine (2.10\,GHz, 187\,GB RAM, HT disabled).
We evaluate three representative programs (NAT, Firewall, and a NIC-resident Memcached cache) and a set of fixed and reconfigurable pipeline architectures (\autoref{fig:details}), including a ``Flex $m\times n$'' family with $m$ stages and $n$ ALUs per stage. We report (i) feasibility/compilation behavior, (ii) synthesis-backed area and timing, and (iii) end-to-end throughput/latency in a full-system environment.

\subsection{Validating End-to-End Evaluation}
A key goal of \sys is to enable early, \emph{trustworthy} system-level evaluation. To validate our methodology, we implement an in-NIC Memcached cache pipeline (akin to NICA~\cite{eran:nica}) and compare full-system simulation results against a physical FPGA testbed. The cache interposes on Memcached UDP traffic and serves GETs directly from NIC state on hit; the host runs unmodified Memcached.
We express the cache as a \sys program (74 lines of DSL), generate a fixed pipeline architecture for it, and generate RTL. The generated pipeline integrates with the open-source Corundum NIC~\cite{forencich:corundum} via an automated flow. In simulation, we plug the NIC into a host running Linux and unmodified Memcached in Gem5, using SimBricks~\cite{li:simbricks} to couple the RTL NIC model with the system. To saturate the server at high hit rates, we attach a packet generator; we synchronize simulator clocks to obtain meaningful throughput/latency measurements.

\begin{figure}%
	\centering%
	\includegraphics[scale=0.40]{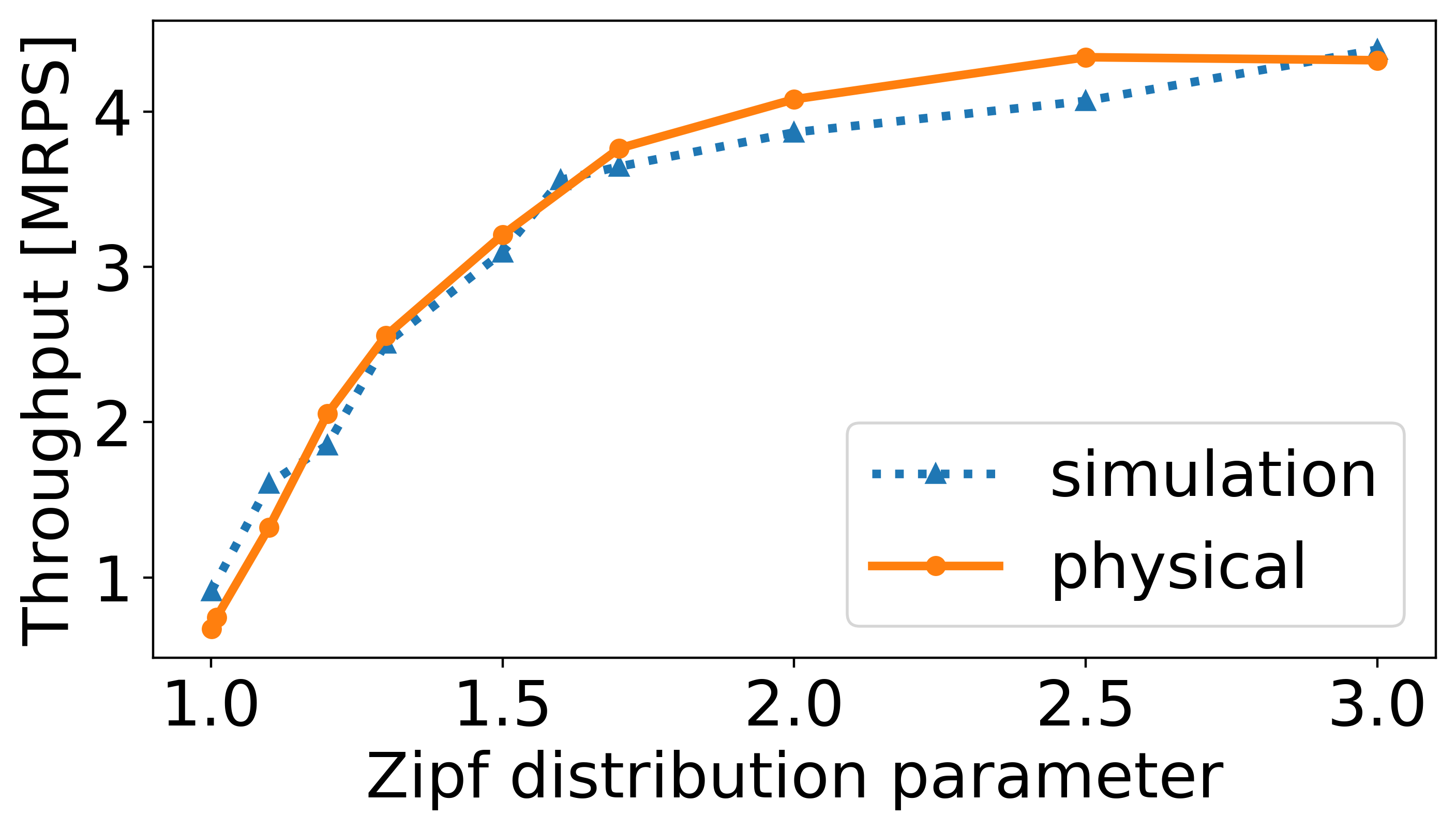}%
	\caption{Memcached-cache throughput vs.\ key-popularity skew: full-system simulation closely matches FPGA measurements.}
	\label{fig:fpga_validation}%
\end{figure}

We synthesize the same RTL for an xcvu9p-flgb2104-2e FPGA (core at 100\,MHz) and evaluate throughput/latency under load across different key-popularity skew levels (\autoref{fig:fpga_validation}). Across the skew range, simulation closely matches the physical testbed with $<10\%$ average relative throughput error. The cache also demonstrates the expected end-to-end impact under skew: the baseline (unmodified Corundum) achieves 180\,kOp/s, while the NIC cache reaches 4.4\,MOp/s with a single client and up to 8.3\,MOp/s at high skew; latency shifts from $>1$\,ms without the cache to the vast majority of requests completing under 100\,$\mu$s with the cache. These results validate that \sys's full-system evaluation methodology yields quantitatively meaningful conclusions.

\subsection{Realism of Generated RTL}
Next, we validate that \sys's RTL generator produces realistic implementations rather than toy artifacts. We replicate an RMT-like baseline from Menshen by extracting architectural details (ALU count, supported instructions, and pipeline structure) and adding explicit state-access modules to match functionality; we remove one metadata-processing ALU that we could not replicate.
This exercise highlights engineering leverage: the Menshen codebase (including optimizations) comprises 9{,}975 lines of Verilog, while our replicated pipeline is described in 160 lines of our architecture DSL. After generating RTL, we synthesize and place-and-route on the same FPGA platform used by Menshen. Our generated design meets the same 250\,MHz timing target. For processing logic, Menshen reports five processing stages at roughly 3{,}000 LUTs each (15{,}948 LUTs total); summing LUT utilization across all ALUs in our design yields 15{,}327 LUTs, comparable to the Menshen processing-stage total. Overall, this indicates that the generator produces timing/area-competitive RTL for non-trivial pipeline architectures.

\subsection{Cost from Synthesis and Failing Proxies}
We use synthesis-backed area and timing to quantify architectural cost and to expose capability--cost trade-offs that proxy metrics miss. \autoref{fig:details} summarizes synthesized area across our fixed baselines and the Flex family. Even simple proxies such as ALU count can be misleading: the fixed Firewall baseline uses 15 ALUs yet synthesizes slightly smaller area than fixed NAT (2.55\,mm$^2$ vs.\ 2.64\,mm$^2$), indicating that non-ALU structure can dominate cost.
Across reconfigurable designs, cost grows non-linearly with scale: Flex~3$\times$100 (304 ALUs) synthesizes to 3.79\,mm$^2$, Flex~10$\times$30 (641 ALUs) to 8.46\,mm$^2$, Flex~30$\times$30 (1{,}921 ALUs) to 15.12\,mm$^2$, and Flex~30$\times$100 (3{,}031 ALUs) to 51.33\,mm$^2$. These results motivate cost-aware exploration grounded in executable hardware rather than stage/ALU-count proxies.

\subsection{Scalability of the Exploration Loop}
We evaluate scalability of hardware generator and compiler.

\begin{figure}
	\centering
	\begin{tabular}{lrrr}
		\toprule
		\textbf{Pipeline Architecture} & \textbf{DSL} & \textbf{RTL} & \textbf{Synthesis} \\
		\midrule
		Fixed NAT (5 ALUs)             & 00:08        & 00:35        & 00:08:50           \\
		Flex $10 \times 30$            & 00:08        & 04:50        & 03:40:08           \\
		Flex $30 \times 100$           & 00:10        & 14:12        & 136:27:02          \\
		\bottomrule
	\end{tabular}
	\caption{Hardware generation runtime breakdown.}
	\label{fig:eval:hwscale}%
\end{figure}

\paragraph{Hardware generation vs.\ synthesis cost.}
We measure three phases: frontend (DSL$\rightarrow$internal abstraction), RTL generation, and external synthesis time (\autoref{fig:eval:hwscale}). Frontend time is negligible. RTL generation scales from 35\,s (fixed NAT) to 14\,min (Flex~30$\times$100), while synthesis dominates end-to-end cost, increasing from 8:50\,min (fixed NAT) to 3:40:08 (Flex~10$\times$30) and 136:27:02 (Flex~30$\times$100). The largest design contains 3{,}031 ALUs and 699k edges. This gap makes it impractical to place full synthesis in the inner loop of exploration and motivates feasibility-first pruning before invoking expensive toolchains.

\begin{figure}%
	\centering%
	\includegraphics[scale=0.37]{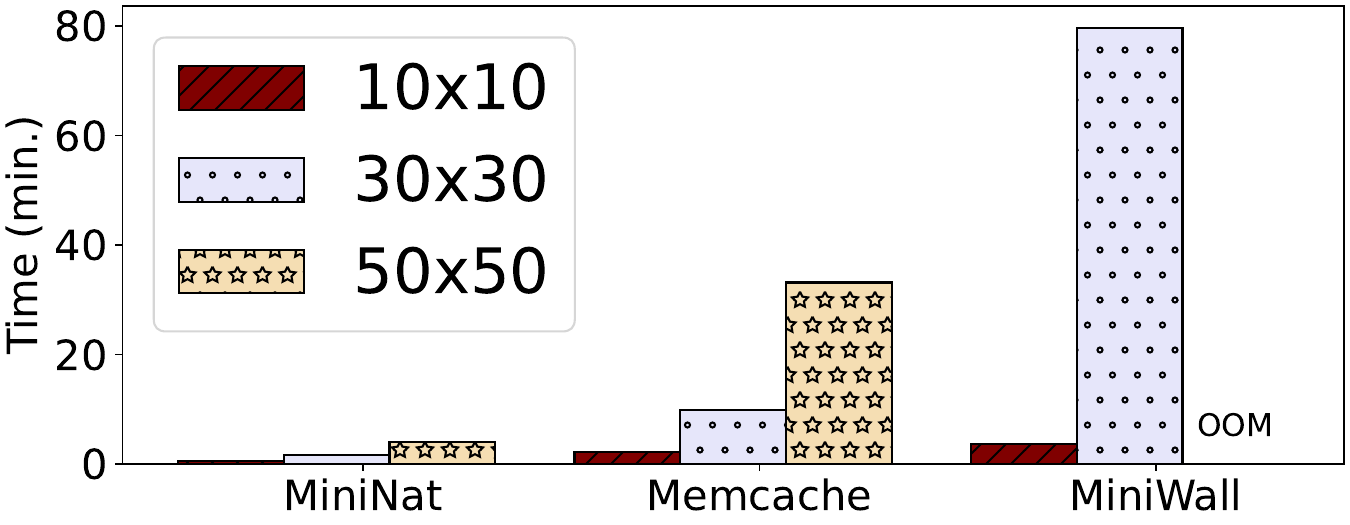}%
	\caption{Compilation time vs.\ pipeline size for three programs with limiter$=10$ (encoding dominates).}
	\label{fig:compiletime}%
\end{figure}

\begin{figure}%
	\centering%
	\includegraphics[scale=0.37]{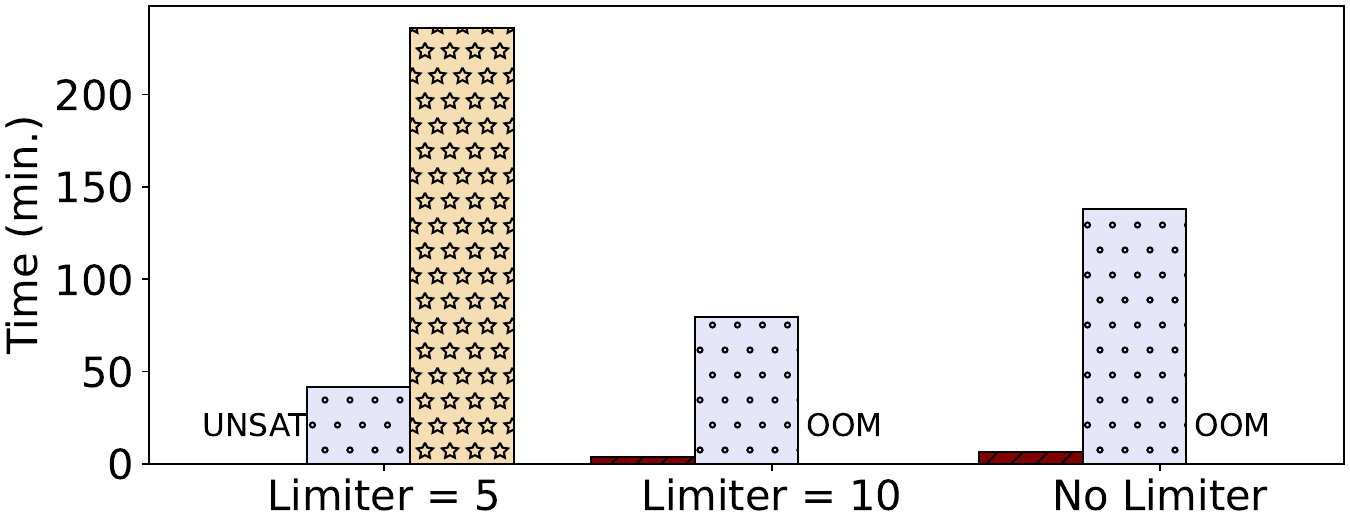}%
	\caption{Limiter sensitivity for MiniWall: smaller limits enable large instances but can over-prune.}
	\label{fig:compilelimiter}%
\end{figure}

\paragraph{Compiler scalability and the limiter trade-off.}
We compile three protocols (MiniNat, Memcache, MiniWall) across 10$\times$10, 30$\times$30, and 50$\times$50 pipeline sizes (\autoref{fig:compiletime}). Compilation is dominated by encoding; SAT solving remains below one minute for all reported successful runs. At the largest size, MiniWall fails with out-of-memory during encoding when using limiter=10, illustrating that encoding scalability is the primary bottleneck. We further study limiter sensitivity on MiniWall (\autoref{fig:compilelimiter}): on 50$\times$50, compilation fails unless the degree limiter is reduced to 5, while on the smallest configuration a limiter of 5 can be overly restrictive and lead to UNSAT. Overall, these results show that the compiler scales to large instances, but that practical exploration benefits from encoding-time pruning mechanisms and explicit scalability controls.

\subsection{Case study: Quantifying Multiplication Support in an RMT-like Pipeline}
Finally, we use a focused case study to demonstrate rapid what-if exploration of capability--cost trade-offs. Starting from an RMT-like pipeline (Flex~3$\times$100 restricted to 3 stages to reduce synthesis time), we add multiplication support to all ALUs via a one-line DSL change. Synthesis shows that multiplication significantly increases cost and worsens timing: area increases from 3.79\,mm$^2$ to 5.34\,mm$^2$, and the design meets timing only at 300\,MHz rather than a 1\,GHz target.
We then mitigate the timing issue by increasing ALU latency from 1 to 3 cycles (again a one-line DSL change), enabling pipelining at the cost of 2 additional cycles per stage. With pipelining, the design meets 1\,GHz timing with area 8.41\,mm$^2$. This case study illustrates the value of executable, synthesis-backed evaluation: small architectural changes can have large, non-linear cost/timing consequences, and \sys enables exploring such trade-offs rapidly and quantitatively.

\subsection{Summary}
Across these experiments, \sys (i) produces trustworthy end-to-end results validated against FPGA measurements, (ii) generates realistic RTL competitive with a hand-imp\-le\-men\-ted baseline, and (iii) scales to large design points, where synthesis is the dominant cost and compilation is dominated by encoding. Together, these results support \sys's feasibility-first, cost-aware, end-to-end exploration methodology.
\section{Discussion}%
\label{sec:discussion}
We clarify when proxy reasoning is sufficient, what our models do (and do not) capture, and how to interpret synthesis-backed cost and full-system performance results.

\subsection{What Proxy Metrics Still Buy You}
Proxy metrics (e.g., stage count, ALUs per stage, aggregate memory) remain useful for coarse filtering and bounding a design space. They often fail once feasibility and cost hinge on \emph{structure}, such as stage boundaries, interconnect topology, and where state access occurs. We view our approach as complementary: proxies narrow candidates, while \sys provides explicit feasibility checking plus synthesis-backed cost and executable end-to-end evaluation for the remainder.

\subsection{Modeling and Front-End Scope}
Our IRs model the structural contract needed for feasibility checking and executable evaluation, not full language semantics or aggressive program rewrites, and not microarchitectural optimizations. These limitations primarily affect \emph{completeness} (a mapping may exist after rewriting) but not \emph{soundness}: any mapping and runtime configuration produced by the compiler yields a correct executable implementation for the modeled pipeline. Extending language support is largely a front-end concern: new sources need only lower to LogicIR, leaving feasibility checking, hardware generation, and evaluation unchanged.

\subsection{Scalable Feasibility-First Compilation}
\sys prioritizes feasibility as an early exploration filter. In practice, scalability is governed mainly by encoding cost rather than SAT solving time, motivating explicit controls such as candidate pruning and degree limiters. These controls preserve \emph{soundness} but can reduce \emph{completeness}; in particular, UNSAT under an aggressive limiter does not imply infeasibility under the full encoding. A practical workflow is to use restrictive settings for fast “find-a-mapping” iterations, then relax controls when definitive infeasibility matters.

\subsection{Interpreting Cost and E2E Results}
We derive cost from synthesis of generated RTL to obtain area and timing for \emph{comparative} exploration; absolute values depend on toolchain, libraries, and backend choices. Nonetheless, synthesis captures structural effects (e.g., interconnect and state-access placement) that dominate relative cost trends across architectures. Our full-system evaluation targets effects that pipeline-only microbenchmarks miss (e.g., backpressure/queueing, DMA behavior, and host-stack interactions) and is validated against FPGA measurements. We recommend reading results in two layers: use synthesis-backed metrics to compare capability--cost trade-offs, and use executable end-to-end evaluation to detect system-level effects that can change architectural rankings.

\subsection{Scope and Generality}
While we focus on programmable packet pipelines, the methodology applies to other staged or spatial accelerators with runtime-configurable structure. Generalizing primarily requires new front ends and architectural block libraries, not a different exploration workflow.
\section{Related Work}%
\label{sec:related}

\paragraph{Programmable packet-processing pipelines and architectures.}
Modern programmable switches and SmartNIC datapaths are commonly organized around the match-action pipeline model popularized by RMT/PISA-style designs and exposed through P4 \cite{bosshart:rmt,bosshart:p4}. A long line of work extends this baseline to improve the capability and performance of stateful and complex workloads, e.g., by adding specialized state/update mechanisms \cite{pontarelli:flowblaze,kim:tea,basat:pint}, by strengthening isolation/multitenancy in shared pipelines \cite{wang:menshen}, or by rethinking the pipeline structure to support deeper stateful dependency patterns and new programming idioms \cite{shrivastav:mp5,feng:empower,chen:optimusprime}. These proposals explore specific architectural points or families, typically evaluated with architecture-specific toolchains and bespoke prototypes. In contrast, \sys targets the \emph{methodology} for exploring \emph{many} candidate pipeline architectures under a common feasibility--cost--system-performance lens, rather than advocating a single new datapath organization.

\paragraph{Dataplane compilers, feasibility checking, and optimization.}
Dataplane compilation has largely focused on mapping programs onto a fixed target pipeline, often with the goal of packing resources or meeting timing constraints. Packet Transactions and related systems provide higher-level programming abstractions for line-rate datapaths and compilation into constrained pipeline-like execution \cite{sivaraman:packettransactions}. More recently, solver-aided compilation has been used to reason about pipeline constraints and minimize required depth for specific targets \cite{gao:cat}. Complementary efforts improve confidence and usability via verification and benchmarking---e.g., p4v for verifying P4 dataplanes and Whippersnapper for standardized evaluation \cite{liu:p4v,dang:whippersnapper}. \sys leverages compilation differently: the compiler acts as a \emph{feasibility oracle} and \emph{configuration generator} across a broad architecture space, enabling systematic pruning and executable evaluation of designs that would otherwise require building (and maintaining) many bespoke compilers.

\paragraph{FPGA-based dataplanes, prototyping, and hardware generation.}
A separate thread compiles dataplane programs into FPGA implementations or uses FPGAs as a rapid prototyping substrate, e.g., P4-to-FPGA and P4-to-NetFPGA workflows \cite{wang:p4fpga,ibanez:p4netfpga}, as well as widely used open platforms and NIC shells \cite{zilberman:netfpgasume,forencich:corundum}. These systems are essential building blocks for deploying and iterating on dataplane hardware, but they typically target either (i) generating hardware specialized to a program, or (ii) prototyping a specific programmable platform. \sys instead automatically generates \emph{programmable pipeline architectures} (as synthesizable RTL) as first-class design points, and uses the same generated hardware both for synthesis-backed cost estimation and for integration into a validated end-to-end evaluation environment.

\paragraph{Design-space exploration and end-to-end simulation.}
Generic design-space exploration (DSE) frameworks and autotuning systems provide powerful optimization infrastructure for navigating large parameter spaces \cite{nardi:practicaldse,ansel:opentuner,paletti:dovado,hsiao:fasttuner}, but they typically treat feasibility constraints and evaluation fidelity as black-box properties of an objective function. Meanwhile, full-system simulation platforms enable faithful end-to-end evaluation of systems and accelerators in realistic software contexts \cite{karandikar:firesim,li:simbricks}. \sys connects these threads for the packet-pipeline domain: it makes feasibility explicit via domain-specific compilation, grounds cost via synthesis of generated RTL, and measures performance via executable end-to-end evaluation, yielding apples-to-apples comparisons across architectural variants.
In this sense, \sys{} is complementary to BO/SMBO-style DSE: it provides an explicit feasibility oracle and an executable evaluator
that such optimizers can call, rather than optimizing proxy metrics in the presence of infeasible points.

\paragraph{In-network and SmartNIC applications.}
Workloads such as in-network caching and key-value acceleration, as well as in-network aggregation for distributed ML, motivate the need for richer dataplane capability and realistic system evaluation \cite{jin:netcache,li:kvdirect,sapio:switchml}. \sys treats these and related applications as representative targets when evaluating architectures, but its contribution is orthogonal to proposing a new in-network service: it provides the tooling to determine which workloads are feasible on which architectures, what capability costs in realizable hardware, and how architectural choices manifest in end-to-end system behavior.
\section{Conclusions}
Designing programmable packet-processing pipelines requires reasoning jointly about program feasibility, hardware cost, and system-level performance---properties that cannot be inferred reliably from proxy metrics. This paper presented \sys, an executable, cost-aware design-space exploration methodology that makes these dimensions explicit and measurable. By decoupling programs from pipeline architectures, generating synthesizable hardware designs, and enabling full-system evaluation, \sys allows designers to identify feasible and effective architectural choices during early design.

Looking forward, \sys provides a foundation for more automated exploration of programmable datapath architectures. Integrating closed-loop search over architectural parameters and extending front-end support for richer programming models are natural next steps. More broadly, we believe that executable, synthesis-backed evaluation will be increasingly important as programmable accelerators become more tightly coupled with system software, and \sys offers a practical path toward that goal.
\section*{Acknowledgments}
\if \ANON 0
	We thank Mehrshad Lotfi for contributing to an early \sys prototype
	implementation.
\fi
We used ChatGPT to revise text, improve flow and correct typos, grammatical errors, and awkward phrasing.
Approach, figures, data, implementation are exclusively human.
Figures have been designed using icons from Flaticon.com.
 
\label{page:last_body}

\bibliographystyle{ACM-Reference-Format}
\bibliography{paper}

\appendix
\section{LogicIR Specification}
\label{appendix:logicir}

\begin{table}[t]%
	\definecolor{tableShade}{gray}{0.9}%
	\centering%
	\begin{tabular}{@{}llp{1.8cm}p{1.3cm}p{1.5cm}@{}}%
		\toprule
		 &              & \textbf{Inputs}          & \textbf{Outputs} & \textbf{Attrs.} \\
		\midrule
		\multicolumn{5}{@{}l@{}}{\textbf{Conversions}}                                  \\
		 & Constant     &                          & value            & value, width    \\
		 & Slice        & operand                  & result           & offset, width   \\
		 & Merge        & a, b, ...                & result           &                 \\
		 & Extend       & operand                  & result           & width, sign     \\
		\multicolumn{5}{@{}l@{}}{\textbf{Compute}}                                      \\
		 & Unary        & operand                  & result           & opcode          \\
		 & Binary       & left, right              & result           & opcode          \\
		 & Conditional  & cond., t-value, f-value  & result           &                 \\
		\multicolumn{5}{@{}l@{}}{\textbf{In / Out}}                                     \\
		 & Packet In    &                          & prefix, length   & prefix len      \\
		 & Packet Out   & cmd, prefix, length      &                  & prefix len      \\
		\multicolumn{5}{@{}l@{}}{\textbf{State}}                                        \\
		 & Array Read   & index                    & value            & array           \\
		 & Array Write  & index, value             &                  & array           \\
		 & Table Lookup & key                      & index            & table           \\
		 & Table Write  & key, \newline index hint & index            & table           \\
		\bottomrule
	\end{tabular}%
	\caption{\sys LogicIR protocol data-flow graph nodes.}%
	\label{tab:dfgnodes}%
\end{table}

This appendix specifies \sys{}'s LogicIR, the architecture-independent
representation of packet-processing programs used for feasibility checking and
for generating an executable runtime configuration. LogicIR intentionally
captures only the \emph{structural} aspects that constrain line-rate execution:
typed data dependencies, explicit state accesses, and fixed bit widths, while
abstracting away stage structure, timing, and microarchitectural details. \sys{}
programs comprise (i) persistent state declarations and (ii) per-packet logic as
a typed data-flow graph (DFG). \autoref{tab:dfgnodes} summarizes the available
DFG node types.

\subsection{Program Structure}

A LogicIR program contains:
\begin{enumerate}
	\item \textbf{State declarations} describing persistent state that survives
	      across packets. LogicIR supports (a) \emph{arrays} (indexed by an integer-like
	      bitvector index) and (b) \emph{tables} (keyed by a fixed-width bitvector key,
	      returning an index used to locate associated per-entry data). State objects
	      are identified by a symbolic name (an identifier) and a fixed shape (widths
	      and capacities).
	\item \textbf{Per-packet logic} expressed as a DFG whose nodes are primitive
	      operations and whose directed edges represent true data dependencies. The DFG
	      contains explicit sources (\texttt{PacketIn}, constants) and sinks
	      (\texttt{PacketOut}, state writes), plus pure compute/conversion nodes.
\end{enumerate}

LogicIR has no loops and no general control-flow graph. Control-dependent value
selection is represented with \texttt{\allowbreak Con\-di\-tio\-nal} nodes (a pure mux), keeping
the representation in a single static DFG.

\subsection{Values, Types, and Bit-Width Discipline}

\paragraph{Bitvectors.}
All values are fixed-width bitvectors. Each node port has a statically known
width. Width-changing operations (\texttt{Slice}, \texttt{Merge}, \texttt{Extend})
make width transformations explicit.

\paragraph{Booleans.}
Predicates are represented as 1-bit values (bitvectors of width 1).

\paragraph{Width consistency.}
Unless otherwise stated by the node type, LogicIR requires that inputs to an
operation have compatible widths. In particular:
\begin{itemize}
	\item \texttt{Binary} compute nodes take same-width operands and produce a
	      result of that width.
	\item \texttt{Conditional} requires \texttt{t-value} and \texttt{f-value} to
	      have equal width; the result has that same width.
	\item State accesses must match the declared state’s value/key widths.
\end{itemize}

\paragraph{Arithmetic semantics.}
LogicIR operations are interpreted over fixed-width bitvectors. When mapped onto
hardware, arithmetic is naturally modulo $2^w$ for width $w$. Signedness is not
a global type property; it is only introduced where needed (e.g., \texttt{Extend}
with a \texttt{sign} attribute).

\subsection{DFG Well-Formedness and Semantics}

A LogicIR DFG is well-formed if:
\begin{itemize}
	\item \textbf{Unique node IDs:} nodes are uniquely identified; each output port
	      is uniquely named within the graph.
	\item \textbf{Port typing:} each input edge connects to an output of identical
	      width.
	\item \textbf{Acyclic data dependencies:} the per-packet DFG is acyclic with
	      respect to pure data edges. Persistent state introduces inter-packet
	      dependence, but not combinational cycles within a packet instance.
	\item \textbf{Explicit state:} every state access node references a declared
	      state object by identifier, and matches its declared shape.
\end{itemize}

\paragraph{Execution meaning.}
LogicIR defines a \emph{combinational} per-packet specification with explicit
state interactions. It does not define an execution order beyond the partial
order induced by edges; any schedule consistent with dependencies is valid.

\subsection{State Declarations}

LogicIR supports two persistent state kinds:
\begin{itemize}
	\item \textbf{Arrays:} declared with \texttt{(id, elem\_width, capacity)}.
	      Reads and writes use \texttt{index} bitvectors. The required index width is a
	      front-end responsibility; the compiler checks that the index can address the
	      declared capacity.
	\item \textbf{Tables:} declared with \texttt{(id, key\_width, capacity)}.
	      Tables map fixed-width keys to indices in a bounded index space (e.g., entry
	      IDs). Table operations are explicit in the DFG and can be used by both
	      dataplane logic (lookups) and dataplane updates (writes).
\end{itemize}

LogicIR does not prescribe coherence policies, eviction, or control-plane
interaction semantics; these are outside the feasibility contract and are
handled by the surrounding system and/or front-end conventions.

\subsection{Node Classes and Per-Node Semantics}

\autoref{tab:dfgnodes} provides the signature-level summary. Below are the key
semantics and invariants.

\paragraph{Conversions.}
\begin{itemize}
	\item \texttt{Constant(width, value)} produces a fixed-width literal.
	\item \texttt{Slice(operand, offset, width)} extracts \texttt{width} bits
	      starting at \texttt{offset}. The result width equals \texttt{width}.
	\item \texttt{Merge(a, b, ...)} concatenates bitvectors in a fixed order
	      (front-end defined), producing a result whose width is the sum of inputs.
	\item \texttt{Extend(operand, width, sign)} extends to \texttt{width} by either
	      zero-extension (\texttt{sign=0}) or sign-extension (\texttt{sign=1}).
\end{itemize}

\paragraph{Compute.}
\begin{itemize}
	\item \texttt{Unary(opcode)} applies a width-preserving unary operation.
	\item \texttt{Binary(opcode)} applies a width-preserving binary operation.
	\item \texttt{Conditional(cond, t, f)} returns $t$ if \texttt{cond} is 1 else
	      $f$ (pure mux).
\end{itemize}
The exact opcode set is a front-end contract parameterized by the hardware
library; candidate generation checks opcode support against the target
architecture’s ALU opcode set.

\paragraph{Packet I/O.}
\begin{itemize}
	\item \texttt{PacketIn(prefix, length)} introduces a value sourced from the
	      packet/metadata namespace. The \texttt{prefix} identifies the region (e.g.,
	      header/metadata selector as used by the front end) and \texttt{length} defines
	      the width in bits.
	\item \texttt{PacketOut(cmd, prefix, length)} is a sink that describes an
	      output action affecting packet/metadata emission. \texttt{cmd} can encode
	      deparser selection/drop semantics as used by the front end; feasibility
	      checking treats \texttt{PacketOut} as an anchored sink that must bind to the
	      corresponding architecture output block.
\end{itemize}

\paragraph{State.}
\begin{itemize}
	\item \texttt{ArrayRead(index, array)} produces the stored value.
	\item \texttt{ArrayWrite(index, value, array)} updates the stored value.
	\item \texttt{TableLookup(key, table)} returns an index for subsequent
	      array accesses (direct-map style).
	\item \texttt{TableWrite(key, index\_hint, table)} returns the index
	      used for the entry (existing or newly allocated, depending on the table
	      implementation policy).
\end{itemize}

\section{ConfigIR Specification}
\label{appendix:configir}

\begin{table}[t]%
	\centering%
	\begin{tabular}{lp{2.7cm}p{2cm}}
		\toprule
		\multirow{2}{*}{\textbf{Name}} & \multicolumn{2}{c}{\textbf{Configuration}}                  \\
		                               & \multicolumn{1}{c}{\textbf{Design-time}}   &
		\multicolumn{1}{c}{\textbf{Run-time}}                                                        \\ \midrule
		\multicolumn{3}{@{}l@{}}{\textbf{Pipeline Elements}}                                         \\
		Register                       & width                                      &                \\[.3em]
		Router                         & inputs                                     & inp. selection \\
		\multicolumn{3}{@{}l@{}}{\textbf{Conversions}}                                               \\
		Constant                       & width                                      & value          \\
		Slice                          & input, offset, width                                        \\
		Merge                          & inputs                                     &                \\
		Extend                         & width, sign                                &                \\[.3em]
		\multicolumn{3}{@{}l@{}}{\textbf{Compute}}                                                   \\
		ALU                            & width, operations, latency                 & op. selection  \\[.3em]
		\multicolumn{3}{@{}l@{}}{\textbf{In / Out}}                                                  \\
		Packet In                      & prefix len, MTU                            &                \\
		Packet Out                     & prefix len, MTU                            &                \\[.3em]
		\multicolumn{3}{@{}l@{}}{\textbf{State}}                                                     \\
		RAM access                     & RAM id, r/w                                &                \\
		CAM access                     & CAM id, r/w                                &                \\
		\bottomrule
	\end{tabular}%
	\caption{\sys ConfigIR building blocks.}%
	\label{tab:hwnodes}%
\end{table}

This appendix specifies \sys{}'s ConfigIR, a structural description of a
reconfigurable packet-processing pipeline architecture. ConfigIR defines the
\emph{execution substrate}: stages, resources, explicit interconnect topology,
and a clear split between design-time parameters (architectural knobs) and
runtime parameters (compiler-assigned configuration). \autoref{tab:hwnodes}
summarizes the ConfigIR element types and their configurable fields.

\subsection{Pipeline Organization and Staging Model}

A ConfigIR pipeline is a directed graph of elements (nodes) connected by typed
edges carrying fixed-width bitvectors. The staged execution model is encoded
explicitly:
\begin{itemize}
	\item \textbf{Stage boundaries} are defined by \texttt{Register} elements.
	      Values written to a stage’s output registers are the only values visible in
	      the next stage.
	\item \textbf{Within-stage logic} is combinational (modulo explicitly modeled
	      multi-cycle blocks via their \texttt{latency} parameter), and must respect the
	      one-stage-per-cycle budgeting assumption used by feasibility checking.
\end{itemize}

ConfigIR may describe multiple pipelines (e.g., ingress / egress) plus input,
output, and storage blocks. Blocks are wired explicitly to pipeline ports so the
compiler can anchor packet I/O and state accesses.

\subsection{Typing and Connectivity Constraints}

\paragraph{Bit-width typing.}
Each element port has a statically known width. Every edge must connect ports of
equal width; width transformations must be expressed explicitly using conversion
elements (constant/slice/merge/extend) or via runtime-configurable conversion
elements when supported.

\paragraph{Structural well-formedness.}
A ConfigIR instance is well-formed if:
\begin{itemize}
	\item All referenced nodes and ports exist and are uniquely identified.
	\item Edge endpoints have matching widths.
	\item Stage boundaries are respected: consumers in stage $i{+}1$ can only see
	      values that pass through the stage-$i$ register boundary.
\end{itemize}

\subsection{Design-Time vs. Run-Time Parameters}

ConfigIR separates:
\begin{itemize}
	\item \textbf{Design-time parameters:} structural knobs fixed when generating
	      RTL (e.g., number of router inputs, ALU width and supported opcode set,
	      memory port structure, ALU latency/pipelining).
	\item \textbf{Run-time parameters:} compiler-produced “machine code” that
	      configures a particular program mapping (e.g., router input selections, ALU
	      opcode selection, constants, conversion parameters, memory IDs and access
	      modes).
\end{itemize}

The compiled runtime configuration assigns all run-time parameters, and is used
unchanged for both synthesized hardware and cycle-accurate simulation.

\subsection{Element Semantics}

\paragraph{Registers.}
\texttt{Register(width)} defines a stage boundary and storage for values that
advance to the next stage. Registers have no runtime parameters.

\paragraph{Routers.}
\texttt{Router(inputs)} models an interconnect selection point. At design time,
\texttt{inputs} fixes the fan-in. At runtime, the compiler sets an \texttt{input
	selection} choosing which predecessor drives the router output for a given
mapping.

\paragraph{Conversions.}
\texttt{Constant}, \texttt{Slice}, \texttt{Merge}, and \texttt{Extend} exist as
explicit architectural resources. Depending on the architecture point, these
may be (i) fully specified at design time, or (ii) configurable at runtime as
summarized in \autoref{tab:hwnodes}. Runtime-configurable conversions let the
compiler realize LogicIR width operations using shared conversion resources.

\paragraph{Compute (ALUs).}
\texttt{ALU(width, operations, latency)} provides a compute resource. The
\texttt{operations} set defines supported opcodes. \texttt{latency} captures
whether the ALU is single-cycle or pipelined. At runtime, the compiler sets the
ALU’s \texttt{op selection} (opcode) and routes operands through the explicit
interconnect graph.

\paragraph{Packet I/O blocks.}
\texttt{PacketIn(prefix len, MTU)} and \texttt{\allowbreak PacketOut(prefix len, MTU)} anchor
the pipeline at the system boundary. They define which packet/metadata slices
are visible and writable, and bound the maximum packet size used by the
generated RTL interface.

\paragraph{State-access blocks.}
ConfigIR models state access explicitly via dedicated access elements. In the
baseline library:
\begin{itemize}
	\item \textbf{RAM access} elements implement array-like indexed storage
	      operations.
	\item \textbf{CAM access} elements implement key-based lookup (table-like)
	      operations.
\end{itemize}
At design time, these elements define port widths and structural placement. At
runtime, the compiler assigns a \texttt{memory id} selecting which concrete
memory instance is accessed, and an \texttt{r/w} mode selecting read vs. write
behavior for that access point.

\subsection{Runtime Configuration Format}

A runtime configuration (the pipeline “machine code”) is a per-element
assignment to all run-time parameters in \autoref{tab:hwnodes}, including:
\begin{itemize}
	\item router input selections,
	\item ALU opcode selections,
	\item constant values,
	\item slice/extend parameters (when these are runtime-con\-fi\-gu\-ra\-ble),
	\item memory IDs and read/write modes for state-access points.
\end{itemize}
This configuration is consumed by the generated RTL through a configuration
interface (e.g., a control register file or configuration bus), and is also used
to drive cycle-accurate simulation.

\section{Feasibility Constraints and SAT Encoding}
\label{appendix:constraints}

This appendix makes the feasibility-first compiler formulation explicit.
The main paper describes feasibility checking as constraint satisfaction over
\emph{placement}, \emph{routing}, and \emph{runtime-parameter selection} (Figure~4 and \S6.2--\S6.5).
Here we define the constraint problem precisely and outline a SAT encoding
sufficient to reimplement \sys's feasibility checker.

\subsection{Inputs, Notation, and Objective}

A LogicIR program is a directed acyclic data-flow graph (DFG)
$G_L = (V_L, E_L)$.
Each LogicIR node $\ell \in V_L$ has:
(i) an operator kind $\mathsf{kind}(\ell)$ (Table~\ref{tab:dfgnodes}),
(ii) a bit-width $\mathsf{w}(\ell)$ for its output value,
and (iii) operator attributes (e.g., opcode for compute, offset/width for slice,
array/table identifier for state access).

Each LogicIR edge $e \in E_L$ is a typed dependency
$e = (\ell_s \rightarrow \ell_t, p)$ from producer $\ell_s$ to consumer input port index $p$
of $\ell_t$. (For unary/binary/conditional/etc., $p$ selects which operand.)

A ConfigIR architecture is a directed graph $G_H = (V_H, E_H)$ whose nodes
$h \in V_H$ are hardware building blocks (Table~\ref{tab:hwnodes}) and whose edges
model interconnect reachability between specific output and input ports.
We write $E_H \subseteq \mathsf{OutPorts}(V_H) \times \mathsf{InPorts}(V_H)$.
Nodes also expose:
\textbf{design-time} parameters (width, supported opcodes, latency, router fan-in, etc.)
and \textbf{runtime} parameters (opcode selections, router selections, constants, slice/extend params,
memory id/mode, etc.).

The compiler solves a \emph{pure feasibility} problem:
\begin{quote}
	Find an assignment to placement, routing, and runtime parameters such that
	the configured ConfigIR datapath is equivalent to the LogicIR DFG for all packets.
\end{quote}
There is no optimization objective (all feasible mappings are equivalent under the
line-rate execution model).

\subsection{Candidate Generation}

Candidate generation computes a finite compatibility relation
$\mathsf{Cand}(\ell) \subseteq V_H$ for each LogicIR node $\ell$.
A hardware node $h \in \mathsf{Cand}(\ell)$ must satisfy, at minimum:
\begin{itemize}
	\item \textbf{Kind compatibility:} $\mathsf{kind}(\ell)$ matches the hardware block class.
	      E.g., $\mathsf{Unary}/\mathsf{Binary}$ map to ALUs; $\mathsf{ArrayRead/Write}$ map to RAM access blocks;
	      $\mathsf{TableLookup/Write}$ map to CAM access blocks; $\mathsf{PacketIn/Out}$ map to pipeline I/O blocks.
	\item \textbf{Width compatibility:} hardware input/output widths can realize $\mathsf{w}(\ell)$, possibly via
	      explicit conversion nodes (Slice / Extend / Merge) when present in LogicIR.
	\item \textbf{Opcode compatibility:} for compute nodes, the ALU's supported opcode set contains the required opcode.
	\item \textbf{State-access compatibility:} for memory nodes, the access block can be configured for the required
	      operation (read vs.\ write, RAM vs.\ CAM) and can reach an appropriate storage block instance.
\end{itemize}

Optionally, candidate generation incorporates the degree limiter (Appendix~\ref{appendix:constraints}.\S\ref{sec:limiter})
by restricting $\mathsf{Cand}(\ell)$ to nodes whose routing neighborhood is within a bounded radius/degree.

\subsection{SAT Variables}

We encode feasibility as SAT over the following variable families.

\paragraph{Placement variables.}
For each LogicIR node $\ell$ and each candidate hardware node $h \in \mathsf{Cand}(\ell)$,
introduce a Boolean variable:
\[
	P_{\ell,h} \equiv \text{``LogicIR node $\ell$ is implemented by HW node $h$''}.
\]

\paragraph{Runtime-configuration variables.}
For each runtime-con\-fi\-gur\-able hardware node $h$, introduce finite-choice variables as one-hot Booleans.
Examples (illustrative; adjust to your concrete ConfigIR schema):
\begin{itemize}
	\item Router $r$ with fan-in $k$: $S_{r,i}$ for $i \in \{0,\dots,k{-}1\}$ where $S_{r,i}$ selects input $i$.
	\item ALU $a$ with opcode set $\mathcal{O}(a)$: $O_{a,op}$ for each $op \in \mathcal{O}(a)$.
	\item Constant block $c$: $V_{c,v}$ for $v$ in the allowed immediate domain (often modeled as a bitvector literal
	      rather than one-hot; if SAT-only, use a Boolean encoding over bits).
	\item RAM/CAM access node $m$ selecting memory instance id from $\mathcal{M}$ and mode in \{R,W\}:
	      $M_{m,id}$ and $RW_{m,\mathsf{R}}/RW_{m,\mathsf{W}}$.
\end{itemize}

\paragraph{Routing / reachability variables.}
We need to connect a producer placement to a consumer input through the \emph{configured} interconnect
(i.e., honoring router selections).
We encode reachability on ports:
for each LogicIR producer node $\ell_s$ and each hardware port $u \in \mathsf{Ports}(V_H)$ in the considered neighborhood,
introduce:
\[
	R_{\ell_s,u} \equiv \text{``Value produced by $\ell_s$ can reach hardware port $u$''}.
\]
In practice, it is sufficient to track reachability to hardware \emph{input ports} of candidate consumer nodes and
to intermediate routing elements (routers/register boundaries) that can forward values without transforming them.

\subsection{Constraints}

The SAT instance is the conjunction of a small number of semantically meaningful constraint classes.

\subsubsection{Placement Correctness}

\paragraph{Exactly-once placement.}
Each LogicIR node is implemented by exactly one compatible hardware node:
\[
	\forall \ell \in V_L:\quad \sum_{h \in \mathsf{Cand}(\ell)} P_{\ell,h} = 1.
\]
Encode equality-to-1 using standard CNF: (i) at-least-one clause over $\{P_{\ell,h}\}$ and
(ii) pairwise (or cardinality-network) at-most-one clauses.

\paragraph{Hardware exclusivity.}
Compute/memory blocks are single-assignment within a configured datapath:
\[
	\forall h \in V_H^{\mathsf{exclusive}}:\quad \sum_{\ell : h \in \mathsf{Cand}(\ell)} P_{\ell,h} \le 1.
\]
Routers and pipeline registers are typically \emph{not} exclusive (they exist to route values), while ALUs and
memory-access nodes typically are.

\subsubsection{Resource Consistency (Runtime Parameters)}

These constraints tie placement decisions to runtime-parameter selections.

\paragraph{ALU opcodes.}
If $\ell$ is a compute node requiring opcode $\mathsf{op}(\ell)$ and $h$ is an ALU candidate,
then placement implies the opcode selection:
\[
	P_{\ell,h} \Rightarrow O_{h,\mathsf{op}(\ell)}.
\]
Additionally, for each ALU $h$ that is used by some $\ell$, enforce exactly-one opcode selection:
\[
	\Big(\sum_{\ell} P_{\ell,h} \ge 1\Big) \Rightarrow \sum_{op \in \mathcal{O}(h)} O_{h,op} = 1.
\]
(You may omit the guard and always enforce one-hot selection for all ALUs if you prefer total configurations.)

\paragraph{Conversion parameters.}
If $\ell$ is Slice/Extend/Constant and $h$ is the matching ConfigIR block, constrain the runtime parameters to match
the LogicIR attributes (offset/width/sign/value). For Boolean SAT, either use one-hot enumerations over legal parameter
tuples or encode bitvector equality with Boolean variables per bit.

\paragraph{Memory id and mode.}
If $\ell$ is an ArrayRead/Write (resp.\ TableLookup/Write) and $h$ is a RAM (resp.\ CAM) access block, then placement
implies:
(i) correct mode (read vs.\ write),
and (ii) a consistent memory-instance identifier selection for the corresponding LogicIR state object.
A simple and effective approach is to introduce a one-hot variable $\mathsf{Bind}_{obj,id}$ per LogicIR state object
(obj) and physical memory id, then require each access to respect that binding:
\[
	P_{\ell,h} \Rightarrow M_{h,id} \quad\text{iff}\quad \mathsf{Bind}_{\mathsf{obj}(\ell),id},
\]
and enforce $\sum_{id} \mathsf{Bind}_{obj,id} = 1$ for each state object.

\subsubsection{Connectivity / Routing Legality}

For each LogicIR edge $e=(\ell_s \rightarrow \ell_t, p)$, if $\ell_s$ is placed at some producer hardware node $h_s$
and $\ell_t$ is placed at some consumer node $h_t$, then the produced value must be routable to the specific input port
$(h_t,p)$, under the configured router selections.

\paragraph{Routing base cases.}
If $\ell_s$ is placed at hardware node $h_s$, then its output port is reachable:
\[
	P_{\ell_s,h_s} \Rightarrow R_{\ell_s,\mathsf{out}(h_s)}.
\]

\paragraph{Propagation across fixed interconnect edges.}
For non-se\-lec\-ting wiring elements (e.g., registers, fixed fanout), propagate reachability along $E_H$:
\[
	\forall (\mathsf{out}(x) \rightarrow \mathsf{in}(y,i)) \in E_H:\quad
	R_{\ell_s,\mathsf{out}(x)} \Rightarrow R_{\ell_s,\mathsf{in}(y,i)}.
\]
Optionally, restrict propagation to routing-only nodes (routers/register boundaries) so that values do not
``flow through'' transforming operators (ALUs/memories) unless that operator is explicitly intended as a passthrough.

\paragraph{Router selection semantics.}
A router $r$ with input ports $\mathsf{in}(r,0..k{-}1)$ and one output $\mathsf{out}(r)$ selects exactly one input:
\[
	\sum_{i=0}^{k-1} S_{r,i} = 1.
\]
Reachability through the router is gated by the selection:
\[
	\forall i:\quad \big(R_{\ell_s,\mathsf{in}(r,i)} \wedge S_{r,i}\big) \Rightarrow R_{\ell_s,\mathsf{out}(r)}.
\]
To avoid ``phantom'' reachability, also constrain the converse direction:
\[
	R_{\ell_s,\mathsf{out}(r)} \Rightarrow \bigvee_{i=0}^{k-1} \big(R_{\ell_s,\mathsf{in}(r,i)} \wedge S_{r,i}\big).
\]

\paragraph{Edge satisfaction.}
Finally, each LogicIR dependency must be satisfied at the consumer input port:
\[
	\begin{gathered}
		\forall (\ell_s \rightarrow \ell_t, p) \in E_L:\quad\\
		\bigvee_{h_s \in \mathsf{Cand}(\ell_s)} \bigvee_{h_t \in \mathsf{Cand}(\ell_t)}
		\Big(P_{\ell_s,h_s} \wedge P_{\ell_t,h_t} \wedge R_{\ell_s,\mathsf{in}(h_t,p)}\Big).
	\end{gathered}
\]

This form is convenient but can be large; in practice, implement it by
introducing helper variables (Tseitin) and constraining only for candidate pairs $(h_s,h_t)$ that are topologically
reachable in $G_H$.

\subsubsection{Packet I/O Anchoring}

Packet ingress/egress nodes must bind to the corresponding architecture blocks.

\paragraph{Ingress.}
For each LogicIR PacketIn node $\ell$, restrict candidates to the architecture's designated PacketIn sources
(and enforce prefix/length compatibility):
\[
	\mathsf{Cand}(\ell) \subseteq V_H^{\mathsf{pktin}}.
\]
If the architecture has multiple packet sources (e.g., ports/queues), additional constraints can bind specific
fields/metadata regions to specific PacketIn instances.

\paragraph{Egress.}
Similarly, PacketOut nodes must map to PacketOut sinks and satisfy deparser command / prefix length constraints.

\subsection{Decoding a Satisfying Assignment}

Given a satisfying assignment, the compiler produces a runtime configuration by:
(i) selecting for each LogicIR node $\ell$ the unique $h$ with $P_{\ell,h}=1$,
(ii) emitting per-node runtime parameters from the corresponding one-hot families
(router selections $S$, ALU opcodes $O$, constants/slice/extend params, memory id/mode),
and (iii) materializing the configured datapath as a per-node ``machine code'' record.
This decoded configuration is used unchanged to drive both synthesized hardware and cycle-accurate simulation.

\subsection{Scalability Heuristics}
\label{sec:limiter}

The bottleneck in large instances is typically encoding size, not SAT solving.
\sys therefore supports \emph{encoding-time pruning} mechanisms that preserve soundness:
any satisfying assignment corresponds to a correct implementation, but pruning may sacrifice completeness.

\paragraph{Degree limiter.}
The (optional) degree limiter restricts the routing neighborhood considered during encoding, e.g., by limiting
the number of hardware edges/nodes explored when building reachability constraints for a candidate placement.
This can dramatically reduce variables/clauses but can yield false UNSAT results:
UNSAT under the limiter does not necessarily imply infeasibility under the full encoding.

\paragraph{Randomized edge subsampling (optional).}
A complementary strategy (used in the older system version) is iterative randomized search:
compile while temporarily ignoring a random subset of hardware interconnect edges.
This never creates invalid solutions (soundness), but may remove all valid routes for a feasible instance, requiring restarts.
Because it cannot prove infeasibility, it should be treated as a best-effort accelerator and paired with the full encoding
when a definitive UNSAT result is needed.

\paragraph{Encoding simplifications.}
Standard CNF engineering applies: introduce Tseitin variables to avoid wide clauses; prefer structured cardinality encodings
(over pairwise) for large one-hot sets; and restrict routing constraints to candidate-relevant subgraphs.
 
\label{page:last}
\end{document}